\documentclass[aps,pre,reprint,10pt,superscriptaddress,showpacs]{revtex4-1}
\usepackage{amsfonts} 
\usepackage{amsmath}
\usepackage{amssymb}
\usepackage{graphicx}
\usepackage{subfigure}
\usepackage{color}
\begin{document}
\title{Effects of Landau damping  on ion-acoustic solitary waves in a semiclassical plasma}
\author{Arnab Barman}
\author{A. P. Misra}
\email{apmisra@visva-bharati.ac.in; apmisra@gmail.com}
\affiliation{Department of Mathematics, Siksha Bhavana, Visva-Bharati University, Santiniketan-731 235, West Bengal, India}
\begin{abstract}
We study the  nonlinear propagation of ion-acoustic waves (IAWs) in an unmagnetized collisionless plasma with the effects of electron and ion  Landau damping  in the weak quantum (semiclassical) regime, i.e., when   the typical  ion-acoustic (IA)   length scale  is larger than the thermal de Broglie wavelength. Starting from a set of classical   and semiclassical Vlasov equations for ions and  electrons,  coupled to the Poisson equation, we derive a modified (by the particle dispersion) Korteweg-de Vries (KdV) equation which governs the evolution of IAWs with the effects of wave-particle resonance.   It is found that in contrast to the classical results,  the nonlinear IAW speed $(\lambda)$ and the linear Landau damping rate $(\gamma)$ are    no longer   constants,  but can vary with the wave number $(k)$ due to the quantum  particle dispersion.  The effects of the    quantum parameter $H$ (the ratio of the plasmon energy  to the thermal energy) and  the electron to ion temperature ratio $(T)$ on the profiles of $\lambda$, $\gamma$ and the solitary wave amplitude are also studied. It is shown that the decay rate of the wave amplitude  is  reduced by the effects of $H$.
 \end{abstract}
\maketitle
\section{Introduction} \label{sec-intro}
Since its theoretical prediction by Landau \cite{landau1946} and later its experimental verification by Malmberg and Wharton \cite{malmberg1964}, the Landau damping has been a longstanding problem of wave-particle interactions in plasmas. The    effects of such collisionless damping  on solitary waves was first theoretically studied  by Ott and Sudan \cite{ott1969}. However, their theory was limited to the effects of electron Landau damping only, and thereby cannot be applied to treat the ion resonance effects. The theory was later developed to include ion Landau damping  adequately and consistently by Vandam \textit{et al.} \cite{vandam1973} by using a multi-scale asymptotic expansion scheme. It has been shown that the nonlinear resonance effects (such as trapping, reflection) can be significant along with the linear resonance of Landau damping.  Though, there have been a  number of similar works in recent times (see, e.g., Refs. \onlinecite{ghosh2011,barman2014,misra2015,chatterjee2015,chatterjee2016,mendonca2016,rightley2016,brodin2015,brodin2016,andreev2016}), however, the theory of Landau damping on nonlinear ion-acoustic waves (IAWs) is still undeveloped in the  semiclassical or full  quantum regimes. 
\par
The purpose of the present work is to extend and generalize the work  of  Vandam \textit{et al.} \cite{vandam1973} in the context of quantum plasmas. Our treatment is, however, limited to the weak quantum regime in which the typical ion-acoustic (IA) length scale is larger than the thermal de Broglie wavelength. We derive a modified Korteweg-de Vries (KdV) equation following the work of Vandam \textit{et al.} \cite{vandam1973}, and  show that even in the weak quantum (semiclassical) regime, the characteristics of  IA solitary waves (IASWs) are significantly modified by the particle dispersion as well as the thermal contributions of electrons and ions.
\section{Basic Equations}  \label{sec:sec1}    
We consider the nonlinear propagation of IAWs in an unmagnetized collisionless   electron-ion semiclassical plasma.  The Vlasov equations for electrons and ions,  respectively, are  \cite{manfredi2001,haas2011}
 \begin{equation}
   \frac{\partial f_e}{\partial t}+v\frac{\partial f_e}{\partial x}+\frac{e}{m_e}\frac{\partial \phi}{\partial x}\frac{\partial f_e}{\partial v}-\frac{e\hbar^2 }{24 m_e^3} \frac{\partial^3 \phi}{\partial x^3} \frac{\partial^3 f_e}{\partial v^3}=0, \label{Vlasov-eqn}
\end{equation}
\begin{equation}
\frac{\partial f_i}{\partial t}+v\frac{\partial f_i}{\partial x}-\frac{e}{m_i}\frac{\partial \phi}{\partial x}\frac{\partial f_i}{\partial v}=0. \label{Vlasov-eqn for ions}
\end{equation} 
Equations \eqref{Vlasov-eqn} and \eqref{Vlasov-eqn for ions} are closed by the Poisson equation 
\begin{equation}
 {\frac{\partial^2 \phi}{\partial x^2}}=-4\pi\sum e_\alpha \int f_\alpha dv. \label{poisson-eq}
\end{equation} 
The semiclassical Vlasov equation \eqref{Vlasov-eqn} is obtained   from the Wigner-Moyal equation with the assumption that  $\hbar/m_ev_{te}L\equiv \lambda_B/L\ll1$ with $\hbar\equiv h/2\pi$ denoting the reduced Planck's constant, $v_{te}$   the electron thermal speed and  $L$   the typical length scale of IAWs.   From Eq. \eqref{Vlasov-eqn},  the classical Vlasov equation can be recovered in the   limit, $\hbar\longrightarrow 0$.  
\par
Next, we normalize the physical quantities according to $v\rightarrow v/c_s$, $\phi\rightarrow e\phi/k_BT_e$, $n_{\alpha}\rightarrow n_{\alpha}/n_{0}$, $f_{\alpha}\rightarrow f_{\alpha}c_s/n_{0}$ where $c_s=\sqrt {k_BT_e/m_i}\equiv\omega_{pi}\lambda_D$ is the IAW speed with $\omega_{pi}=\sqrt{4\pi n_{0}e^2/m_i}$ and $\lambda_D=\sqrt{k_BT_e/4\pi n_{0}e^2}$ denoting, respectively, the ion plasma frequency and the plasma Debye length. Furthermore,  $n_{0}$ is the equilibrium number density of electrons and ions, $T_j$ is the thermodynamic temperature of electrons $(j=e)$ or ions $(j=i)$, $k_B$ is the Boltzmann constant. The space and time variables are normalized by $\lambda_D$ and $\omega^{-1}_{pi}$ respectively. 
 \par  
   Thus, from Eqs. \eqref{Vlasov-eqn}-\eqref{poisson-eq}, we obtain the following set of equations in dimensionless   forms: 
\begin{equation}
 \frac{\partial f_e}{\partial t}+v\frac{\partial f_e}{\partial x}+\frac{1}{m}\frac{\partial \phi}{\partial x}\frac{\partial f_e}{\partial v}-\frac{H^2}{24m^2}  \frac{\partial^3 \phi}{\partial x^3} \frac{\partial^3 f_e}{\partial v^3}=0, \label{nond-Vlasov-eqn-electron} 
\end{equation}
\begin{equation}
 \frac{\partial f_i}{\partial t}+v\frac{\partial f_i}{\partial x}-\frac{\partial \phi}{\partial x}\frac{\partial f_i}{\partial v}=0, \label{nond-Vlasov-eqn-ion} 
\end{equation}
\begin{equation}
 {\frac{\partial^2 \phi}{\partial x^2}}=-\sum_{\alpha=e,i} \theta_{\alpha} \int f_\alpha dv, \label{nond-poisson-eq}
\end{equation} 
 where $m=m_e/m_i$ is the electron to ion mass ratio,  $H=\hbar\omega_{pe}/k_BT_e$   is the dimensionless  quantum parameter  denoting the ratio of the electron plasmon energy to the thermal energy and $\theta_\alpha=\mp1$ for electrons $(\alpha=e)$  and ions $(\alpha=i)$. 
  \section{Derivation of KdV equation} \label{sec:sec2}
  In this section, we derive the evolution equation for small  amplitude IAWs  following the same multi-scale asymptotic expansion  technique as in Ref. \onlinecite{vandam1973} in which the  physical quantities are perturbed from the equilibrium state as  
\begin{equation}
\begin{split}
\phi=&\epsilon\phi^{(1)}+\epsilon^2\phi^{(2)}+\cdots, \\
f_{\alpha}=&f^{(0)}_{\alpha}+\epsilon f^{(1)}_{\alpha}+\epsilon^2 f^{(2)}_{\alpha}+\cdots,\label{expansions}
\end{split}
\end{equation} 
where $\epsilon~(\lesssim1)$ is a small positive scaling parameter  measuring the weakness of perturbations. The equilibrium distribution of electrons and ions, i.e.,  $f^{(0)}_{\alpha}$, for $\alpha=e, i$, are assumed to be the Maxwellian given by 
\begin{equation}
f^{(0)}_{\alpha}=\frac{1}{\sqrt{{2\pi}}}\sqrt{\frac{m_{\alpha}T_e}{m_iT_{\alpha}}}\text{exp}\left(-\frac{m_{\alpha}T_e}{m_iT_{\alpha}}\frac{v^2}{2}\right).\label{f_alpha Maxwellian distribution}
\end{equation} 
Such an assumption for electrons is valid when $T_e\gg T_F$, where $T_F$ is the electron Fermi temperature. Also, the  expansion  for $f_{\alpha}$ in \eqref{expansions} is valid only in the non-resonance region where $|v-\lambda|\gg o(\epsilon)$ with $\lambda$ denoting the phase velocity of the wave. However, in the resonance region, to be discussed later, where $v\approx\lambda$, a slightly different ordering of $\epsilon$ is to be considered. Furthermore,  in order to properly include  the contributions of resonant   particles (trapped and/or free)   (Note here that the     Gardner-Morikawa transformation \cite{vandam1973,washini1966} 
$\zeta=\epsilon^{1/2}(x-\lambda t)$, $s=\epsilon^{3/2}x$, usually used in fluid models, can not be applied directly  to the Vlasov equation), we introduce the stretched coordinates and the multi-scale Fourier-Laplace transforms for $f_{\alpha}-f^{(0)}_{\alpha}$ and $\phi$ as \cite{vandam1973} 
\begin{equation}
\xi=\epsilon^{1/2}x,~ \sigma=\epsilon^{1/2}t,~s=\epsilon^{3/2}x, \label{stretching}
\end{equation}
\begin{equation}
 \begin{split}
 f^{(n)}_{\alpha}(v,\xi,\sigma,s)=& \frac {i}{(2\pi)^2}\int_ {-\infty}^{\infty}dk \int_W d\omega \hat{f}^{(n)}_{\alpha}(v,k,\omega,s)\\
 &\times(\omega-k\lambda)^{-1}\exp[i(k\xi-\omega\sigma)], \\
 \phi^{(n)}(\xi,\sigma,s)= &\frac {i}{(2\pi)^2} \int_{-\infty}^{\infty}dk\int_W d\omega \hat{\phi}^{(n)} (k,\omega,s)\\
 &\times(\omega-k\lambda)^{-1}\exp[i(k\xi-\omega\sigma)], \label{Fourier-Lap-int}
\end{split}
 \end{equation}
  where  $\omega~(k)$ is the wave frequency (number),  $W$ is the usual Laplace transform contour lying above any poles of the integrals on the complex $\omega$-plane, and the transformed quantities $\hat{f}^{(n)}_{\alpha}$ and $\hat{\phi}^{(n)}$ are analytic for all $\omega$, i.e.,   the inverse transforms does not contain any factor $1/\left(\omega-\lambda k\right)$ in their expressions.  Next, we substitute   Eqs. \eqref{expansions} and \eqref{stretching} into Eqs. \eqref{nond-Vlasov-eqn-electron}-\eqref{nond-poisson-eq}, and equate different powers of $\epsilon$.  The results are obtained in Secs. \ref{sec-nonresonance-1st-order} and \ref{sec-nonresonance-2nd-order}.
\subsection{First-order perturbations and nonlinear wave speed} \label{sec-nonresonance-1st-order}
 Equating the coefficients of $\epsilon^{3/2}$ from  Eqs. \eqref{nond-Vlasov-eqn-electron} and \eqref{nond-Vlasov-eqn-ion}, and the coefficients  of $\epsilon$ from Eq. \eqref{nond-poisson-eq}, we, respectively, obtain 
 \begin{eqnarray}
  &&\frac{\partial f^{(1)}_\alpha}{\partial\sigma}+v\frac{\partial f^{(1)}_\alpha}{\partial\xi}-\theta_\alpha\left(\frac{m_i}{m_\alpha}\right)G_{\alpha}(v)\frac{\partial\phi^{(1)}}{\partial\xi}\notag\\
  &&+R_\alpha\frac{\partial^2G_{\alpha}(v)}{\partial v^2}\frac{\partial^3\phi^{(1)}}{\partial\xi^3}=0,\label{f_alpha-1}
 \end{eqnarray}
   \begin{equation}
 \sum_{\alpha=e,i}\theta_{\alpha}\int_{\text{n.r.}}f^{(1)}_{\alpha}dv=0,\label{1st order poisson}
 \end{equation}
 where   \textit{n.r.} stands for the non-resonance region in the velocity integral,   $G_{\alpha}(v)=\partial f^{(0)}_{\alpha}/\partial v$,     $R_e=-H^2/24m^2$ for quantum electrons and  $R_i=0$ for classical ions. Also,    we assume that $R_e(\partial^3f_e^{(0)}/\partial v^3)\sim O(\epsilon^{-1})$ in the non-resonance region in order to include the contribution from the particle dispersion in the linear and nonlinear regimes. Note that, the quantum effect  of electrons gives rise a new term proportional to $R_e$, and  Eq. \eqref{1st order poisson} gives the quasineutrality condition, i.e., the electron and ion number densities  do not  differ  until  $o(\epsilon^2)$. The resonance effects such as the Landau damping    will be shown to be at least of $o(\epsilon^2)$. 
 \par
\begin{figure*}
\includegraphics[scale=0.5]{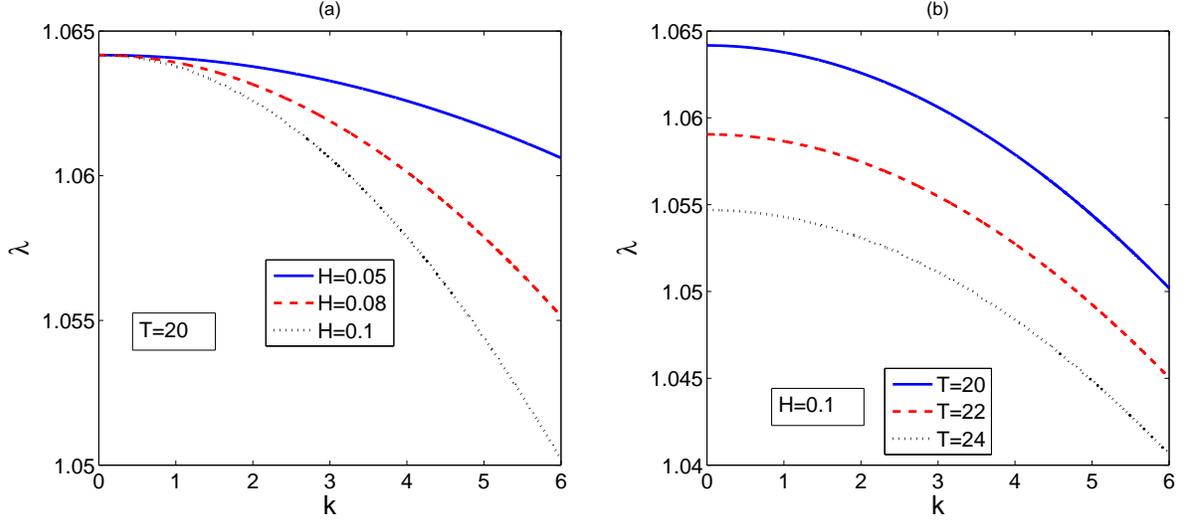}
\caption{Plot of the phase velocity $\lambda$ (normalized by $c_s$), given by Eq. \eqref{lambda1}, is shown against the wave number $k$ (normalized by $\lambda_{D}^{-1}$) in two different cases: (a) when $T$ is fixed and $H$ varies (b) when $H$ is fixed and $T$ varies. }
\label{fig:phase-velo}
\end{figure*}

Transforming Eq. \eqref{f_alpha-1} according to the formula \eqref{Fourier-Lap-int} and   using the fact that $f_\alpha (x,v,t)=f_\alpha^{(0)}(v)$ and $\phi(x,t)=0$ for $t<0$, we obtain
 \begin{equation}
 \hat f^{(1)}_{\alpha}=-\frac{k}{\omega-kv}\left(\theta_{\alpha}\frac{m_i}{m_{\alpha}}G_{\alpha}(v)+k^2R_\alpha\frac{\partial^2G_{\alpha}(v)}{\partial v^2}\right) \hat\phi^{(1)}.\label{hat f1}
 \end{equation}
 Substituting the expression for $f^{(1)}_{\alpha}$, obtained from the  Fourier-Laplace inversion of Eq. \eqref{hat f1}, into Eq. \eqref{1st order poisson}, we obtain the following dispersion relation for the nonlinear wave speed $\lambda\equiv\omega/k$  
 \begin{equation}
 \sum_{\alpha=e,i}\int_\text{n.r.}\frac{1}{v-\lambda}\left(\frac{m_i}{m_{\alpha}}G_{\alpha}(v)+\theta_{\alpha} k^2R_\alpha\frac{\partial^2G_{\alpha}(v)}{\partial v^2}\right)dv=0. \label{dispersion relation}
 \end{equation}
 Note that the dispersion relation is modified by the quantum correction proportional to $R_\alpha$, in absence of which we can recover the classical results as in Ref. \onlinecite{vandam1973}. 
Using the unperturbed velocity distribution functions for electrons and ions given by Eq. \eqref{f_alpha Maxwellian distribution}, and appropriate limiting expressions for the Plasma dispersion function we obtain (for details, see Appendix \ref{appendix-dispersion}) the following dispersion relation for IAWs in semiclassical plasmas:
  \begin{equation}
\lambda^2=\frac{1+\sqrt{1+\frac{12}{T}\left(1+\frac{H^2k^2}{12}\right)}}{2\left(1+\frac{H^2k^2}{12}\right)}.\label{lambda1}
\end{equation}
In the limit of $T\equiv T_e/T_i\gg1$ and   $H^2k^2/12\ll1$, the expression for $\lambda$ in  Eq. \eqref{lambda1} can be approximated as  
\begin{equation}
\lambda\approx1+\frac{3}{2T}-\frac{H^2k^2}{24}. \label{lambda2}
\end{equation}
In terms of the original dimensional variables, Eq. \eqref{lambda2} is rewritten as
\begin{equation}
\lambda=c_s\left(1+\frac{3}{2}\frac{T_i}{T_e}-\frac{H^2}{24}k^2\lambda_D^2\right). \label{lambda3}
\end{equation}
 The expression \eqref{lambda2} of $\lambda$   exactly agrees with Eq. (2.37) of Ref. \onlinecite{vandam1973}   in the formal semiclassical limit of $H=0$.  Thus, in contrast to the quantum fluid theory \cite{haas2003} or classical kinetic theory \cite{vandam1973}, the phase velocity of the nonlinear IAWs, in presence of the particle's dispersion, may no longer be a constant but can vary with the wave number $k$ and gets modified by the quantum parameter $H$. The profiles of $\lambda$  for different values of  the quantum parameter $H$ and the temperature ratio $T$ are shown in Fig. \ref{fig:phase-velo}. It is seen that the wave speed $\lambda$     decreases with the wave number, and this diminution is greatly enhanced  with a small increase of the quantum parameter $H$ [Fig. \ref{fig:phase-velo}(a)]. Furthermore, a significant reduction in $\lambda$ is also seen to occur  with an enhancement of the   temperature ratio $T$ [Fig. \ref{fig:phase-velo}(b)]. From Eq. \eqref{lambda2}, one can  obtain a value of the phase velocity for a given set of plasma parameters. For example, for typical parameter values with $T=20$, $H=0.05$ (which, e.g.,  correspond to plasmas with $T_e\sim7\times10^6$ K, $n_0\sim6\times10^{23}$ cm$^{-3}$, so that $\lambda_D=3\times10^{-9}$ cm) and given a dimensionless wave number $k=2$ (i.e., in dimensional variable, $k=6.5\times10^8$ cm$^{-1}$), one obtains the   phase velocity $\lambda=1.06$ or in dimensional variable, $\lambda=9.6\times10^7$ cm/s.   We mention that further increase of the values of $H$ and $k$ may not be admissible as we are interested in the propagation of IAWs in the weak quantum regime $H<1$. However, the values of  $T$ may be considered in the range $20\lesssim T\lesssim100$. Furthermore, in the   range of values of $k$ (as in Fig. \ref{fig:phase-velo}), $\lambda$  is greater than the unity, implying that the relation $v_{ti}\ll\lambda\ll v_{te}$  holds good for IAWs with $T\gg1$ and $m\equiv m_e/m_i\ll1$. Here,  $v_{t\alpha}=\sqrt{k_BT_\alpha/m_\alpha}$ is the thermal velocity of   $\alpha$-spices particle. As expected, in the limit of $k\rightarrow0$ and/or in the semiclassical limit $H\rightarrow0$, the phase velocity of IAWs approaches a constant value, i.e., close to the ion-acoustic speed $c_s$.
\subsection{Second-order perturbations and KdV equation} \label{sec-nonresonance-2nd-order}
 Equating the coefficients of $\epsilon^{5/2}$ from  Eqs. \eqref{nond-Vlasov-eqn-electron} and \eqref{nond-Vlasov-eqn-ion}, and the coefficient of $\epsilon^2$ from Eq. \eqref{nond-poisson-eq}, we respectively, obtain 
\begin{eqnarray}
&&\frac{\partial f^{(2)}_{\alpha}}{\partial\sigma}+v\left(\frac{\partial f^{(2)}_{\alpha}}{\partial\xi}+\frac{\partial f^{(1)}_{\alpha}}{\partial s}\right)-\theta_{\alpha}\left(\frac{m_i}{m_\alpha}\right)\notag\\
&&\times\left[G_{\alpha}(v)\left(\frac{\partial\phi^{(2)}}{\partial\xi}+\frac{\partial\phi^{(1)}}{\partial s}\right)+\frac{\partial\phi^{(1)}}{\partial\xi}\frac{\partial f^{(1)}_\alpha}{\partial v}\right]\notag\\
&&+R_\alpha\left[\frac{\partial^3\phi^{(2)}}{\partial\xi^3}+3\frac{\partial}{\partial s}\left(\frac{\partial^2\phi^{(1)}}{\partial\xi^2}\right)\right]\frac{\partial^3 f^{(0)}_\alpha}{\partial v^3}=0, \label{2nd order Vlasov equation}
\end{eqnarray}
and
\begin{equation}
\frac{\partial^2\phi^{(1)}}{\partial\xi^2}=-\sum\theta_\alpha\left(\int_\text{n.r.}f^{(2)}_\alpha dv+\left< \int_\text{res.}f_{\alpha}dv\right>^{(2)}\right),\label{2nd order poisson}
\end{equation}
where  the   term in the angular brackets    appears due to  the second order density perturbation in the resonance region. Transforming Eq. \eqref{2nd order Vlasov equation} according to the formula \eqref{Fourier-Lap-int} and eliminating $\hat f^{(1)}_\alpha$ by using Eq. \eqref{hat f1}, we obtain
\begin{eqnarray}
&&\hat f^{(2)}_{\alpha}=\left[-k\left(\theta_\alpha\left(\frac{m_i}{m_\alpha}\right)G_\alpha (v)+R_\alpha k^2\frac{\partial^2 G_\alpha(v)}{\partial v^2}\right)\hat \phi^{(2)}\right.\notag \\
&&\left.+\left(\theta_\alpha\left(\frac{m_i}{m_\alpha}\right)\text{i}\omega(\omega-kv)^{-1} G_\alpha (v)+3\text{i}R_\alpha k^2\frac{\partial^2 G_\alpha(v)}{\partial v^2}\right)\right.\notag\\
&&\left.\times\frac{\partial\hat \phi^{(1)}}{\partial s}+\text{i}\theta_\alpha\left(\frac{m_i}{m_\alpha}\right)(\omega-k\lambda)\tilde{B}_\alpha\right](\omega-kv)^{-1}, \label{hat f2}
\end{eqnarray}
where
\begin{equation}
\tilde{B}_\alpha=-\text{i}\int_{-\infty}^{\infty}d\xi\int_{-\infty}^{\infty}d\sigma\frac{\partial\phi^{(1)}}{\partial\xi}\frac{\partial f^{(1)}_\alpha}{\partial v}\text{exp}[-\text{i}(k\xi-\omega\sigma)]\label{tilde B_alpha}
\end{equation}
is the transform of the nonlinear term $(\partial\phi^{(1)}/\partial\xi)(\partial f^{(1)}_\alpha/\partial v)$. Next, we substitute the expressions of $f^{(1)}_\alpha$ and $\phi^{(1)}$ from Eq. \eqref{Fourier-Lap-int} into Eq. \eqref{tilde B_alpha}.  By virtue of Eq. \eqref{hat f1} and using the Convolution theorem,  Eq. \eqref{tilde B_alpha} reduces to
\begin{equation}
\begin{split}
\tilde{B}_\alpha (\omega, k)=&\frac{\theta_\alpha}{(2\pi)^{2}}\left(\frac{m_i}{m_\alpha}\right)\int\int_{W'}k'(k-k')(\omega'-\lambda k')^{-1}\\
&\times[(\omega-\omega')-\lambda(k-k')]^{-1}\hat\phi^{(1)}(\omega', k')\\
&\times\hat\phi^{(1)}(\omega-\omega', k-k')\frac{\partial}{\partial v}\left[\frac{G_{\alpha}(v)}{\omega'-vk'}\right]d\omega'dk', \label{tilde B_alpha1}
\end{split}
\end{equation}
where $\text{Im} (\omega-\omega')>0$.
Taking the Fourier inversion of Eq. \eqref{hat f2} and integrating both sides of this equation over the non-resonance region, we obtain
\begin{equation}
\begin{split}
\int_{\text{n.r.}}f^{(2)}_{\alpha}dv=&\theta_\alpha\left(\frac{m_i}{m_\alpha}\right)\left[\phi^{(2)}(\zeta, s)\int_{\text{n.r.}}(v-\lambda)^{-1}G_\alpha (v)dv\right. \\
&\left.-A_\alpha-\theta_\alpha B_\alpha\right]+R_\alpha k^2\phi^{(2)}\int_{\text{n.r.}}(v-\lambda)^{-1}\\
&\times\frac{\partial^2 G_\alpha(v)}{\partial v^2}dv, \label{f2}
\end{split}
\end{equation}
where  $\zeta=\xi-\lambda\sigma$ and $A_\alpha$,  $B_\alpha$ are given by
\begin{equation}
\begin{split}
A_\alpha=&(2\pi)^{-2}\int\int_{W}\frac{dkd\omega}{\omega-\lambda k}\frac{\partial\hat\phi^{(1)}}{\partial s}\exp{[ i(k\xi-\omega\sigma)]}\\
&\times\int_{\text{n.r.}}\left(\omega G_\alpha(v)+3\theta_\alpha R_\alpha k^2\frac{\partial^2 G_\alpha(v)}{\partial v^2}\right)\frac{dv}{(\omega-kv)^{2}},\label{A_alpha}
\end{split}
\end{equation}
\begin{equation}
\begin{split}
B_\alpha=&(2\pi)^{-4}\int\int_{W}dkd\omega\int\int_{W'}dk'd\omega'k'(k-k')\\
&\times(\omega'-k'\lambda)^{-1}[(\omega-\omega')-\lambda(k-k')]^{-1}\\
&\times\hat\phi^{(1)}(\omega', k')\hat\phi^{(1)}(\omega-\omega', k-k')\text{exp}[\text{i}(k\xi-\omega\sigma)]\\
&\times L_\alpha(\omega, k; \omega', k'),\label{B_alpha}
\end{split}
\end{equation}
with  
\begin{equation}
L_\alpha=-k\int_{\text{n.r.}}(\omega'-vk')^{-1}(\omega-vk)^{-2} G_\alpha(v)dv, \label{L_alpha}
\end{equation}
and to the fact that  over the non-resonance region, where $|v-\lambda|>>o(\epsilon)$,  $G_{\alpha}(v),~(\omega-kv)^{-1},~(\omega'-k'v)^{-1}\longrightarrow 0$. 
Closing the $W$ contour from below and differentiating Eq. \eqref{A_alpha} with respect to  the transformation $\zeta=\xi-\lambda\sigma$, we obtain
\begin{equation}
\begin{split}
\frac{\partial A_\alpha}{\partial\zeta}=&k\frac{\partial\phi^{(1)}}{\partial s}\int_{\text{n.r.}}\left(\omega G_\alpha(v)+3\theta_\alpha R_\alpha k^2\frac{\partial^2 G_\alpha(v)}{\partial v^2}\right)\\
&\times (\omega-kv)^{-2}dv\\
=&\left[A_{0\alpha}(\lambda)+C_{0\alpha}(\lambda)\right]\frac{\partial\phi^{(1)}}{\partial s}, \label{differentiation of A_alpha}
\end{split}
\end{equation}
where, by virtue of Eq. \eqref{dispersion relation},  $A_{0\alpha}$ and  $C_{0\alpha}$  are given as
\begin{equation}
\sum_\alpha  A_{0\alpha}(\lambda)=\lambda\sum_\alpha \left(\frac{m_i}{m_\alpha}\right)\int_{\text{n.r.}}\frac{dv}{v-\lambda}\frac{\partial G_\alpha(v)}{\partial v}, \label{A_0alpha}
\end{equation}
\begin{equation}
\sum_\alpha C_{0\alpha}(\lambda)=18k\sum_\alpha \left(\frac{m_i}{m_\alpha}\right)\theta_\alpha R_\alpha\int_{\text{n.r.}}\frac{G_\alpha(v)}{(v-\lambda)^4} dv, \label{C_0alpha}
\end{equation}
and we have used the fact that  over the non-resonance region, $(v-\lambda)^{-1}$, $(v-\lambda)^{-2}$ and $(v-\lambda)^{-3}\longrightarrow 0$.
In the expression of $B_\alpha$ [Eq. \eqref{B_alpha}],    the  $\omega'$-integration is performed with the fact that the pole $\omega'=\omega-\lambda(k-k')$ lies on the $W$ contour which is above the $W'$ contour, and we close the $W'$ contour in the lower half plane, so that only the residue at  $\omega'=\lambda k'$ contributes. However, the $\omega$-integration is performed by closing the $W$ contour from below and assuming the pole is at $\omega=\lambda k$.    Thus,  we obtain
\begin{equation}      
\begin{split}
B_\alpha=&-B_{0\alpha}(\lambda)(2\pi)^{-2}\int\int_{\bar{W'}}k^{-1}(k-k')\hat\phi^{(1)}(k-k')\\
&\times\hat\phi^{(1)}(k')\text{exp}(\text{i}k\zeta)dkdk',\label{transformed B_alpha}
\end{split}
\end{equation}
where  we have used
\begin{equation}
\begin{split}
L_\alpha(k, k')=&-k\int_{\text{n.r.}}\frac{1}{k^2k'}\left(\frac{\omega'}{k'}-v\right)^{-1}\left(\frac{\omega}{k}-v\right)^{-2}G_\alpha dv,\\
=&\frac{1}{kk'}\int_{\text{n.r.}}(v-\lambda)^{-3}G_\alpha(v) dv, \label{L_alpha(k,k')}
\end{split}
\end{equation}
and so, $B_{0\alpha}(\lambda)$ is given by
\begin{equation}
B_{0\alpha}(\lambda)=\int_{\text{n.r.}}(v-\lambda)^{-3}G_\alpha(v) dv. \label{B_0alpha}
\end{equation}
Next, differentiating Eq. \eqref{transformed B_alpha} with respect to $\zeta$  and using the convolution theorem, we obtain
\begin{equation}
\frac{\partial B_\alpha}{\partial\zeta}=-2B_{0\alpha}(\lambda)\phi^{(1)}\frac{\partial\phi^{(1)}}{\partial\zeta}, \label{differentiation of B_alpha}
\end{equation}
Finally, differentiating Eq. \eqref{2nd order poisson}   with respect to $\zeta$ and using Eq. \eqref{f2},  
  we note that  the coefficient of  $\partial\phi^{(2)}/\partial\zeta$ vanishes by   the dispersion relation \eqref{dispersion relation}. Thus, we obtain the following modified KdV equation 
\begin{equation}
\begin{split}
\frac{\partial^3\phi^{(1)}}{\partial\zeta^3}&+a\frac{\partial\phi^{(1)}}{\partial s}+b\phi^{(1)}\frac{\partial\phi^{(1)}}{\partial\zeta}\\
&+\frac{\partial}{\partial\zeta}\sum_\alpha\theta_\alpha\left<\int_{\text{res.}}f_\alpha dv\right>^{(2)} 
=0, \label{modified KdV}
\end{split}
\end{equation} 
where     the coefficients $a$ and  $b$ are given by
\begin{equation}
a=-\sum_\alpha\left(\frac{m_i}{m_\alpha}\right)[A_{0\alpha}(\lambda)+C_{0\alpha}(\lambda)], \label{a}
\end{equation}
and
\begin{equation}
b=\sum_\alpha\theta_\alpha\left(\frac{m_i}{m_\alpha}\right)\int_{\text{n.r.}}(v-\lambda)^{-3}G_\alpha(v)dv. \label{b}
\end{equation}
 \par
In order to evaluate the integral in Eq. \eqref{modified KdV} over the resonance region where the particle velocity approaches the phase velocity of the wave, a different ordering for the distribution function is to be considered. Here, we assume that  $v-\lambda\sim\epsilon^{1/2}(m_i/m_\alpha)^{1/2}u$, where $u\sim o(1)$, and   along the particle path $f_\alpha(v, x, t)=f_\alpha(v, 0, 0)=f^{(0)}_\alpha(v)$. So, in the resonance region, we expand the  distribution function  as \cite{vandam1973}
\begin{equation}
f_\alpha=f^{(0)}_\alpha+\epsilon^{3/2}\left(\frac{m_\alpha}{m_i}\right)^{1/2}f_\alpha^{(1)}+\cdots,\label{expansion in resonance-region}
\end{equation}
together with the ordering for the derivatives
\begin{equation}
\frac{\partial f^{(0)}_\alpha}{\partial v}\left(\frac{m_i}{m_\alpha}\right)\sim o(\epsilon),~\frac{\partial f^{(1)}_\alpha}{\partial v}\left(\frac{m_i}{m_\alpha}\right)^{1/2}\sim o(\epsilon^{-1/2}).\label{orderings}
\end{equation}
The expansion for $\phi$ remains the same as in Eq. \eqref{expansions}. Thus, from Eqs. \eqref{nond-Vlasov-eqn-electron} and \eqref{nond-Vlasov-eqn-ion},  the coefficients  of $\epsilon^{5/2}$ yield 
\begin{eqnarray}
&&\epsilon^{-1/2}\left(\frac{m_\alpha}{m_i}\right)^{1/2}\left[\frac{\partial f^{(1)}_\alpha}{\partial\sigma}+v\frac{\partial f^{(1)}_\alpha}{\partial\xi}\right]-\theta_\alpha\left[\left(\frac{m_i}{m_\alpha}\right)\epsilon^{-1}\right.\notag\\
&&\left.\times\frac{\partial f^{(0)}_\alpha}{\partial v}\right]\frac{\partial\phi^{(1)}}{\partial\xi}-\theta_\alpha\left[\left(\frac{m_i}{m_\alpha}\right)^{1/2}\frac{\partial f^{(1)}_\alpha}{\partial v}\epsilon^{1/2}\right]\frac{\partial\phi^{(1)}}{\partial\xi}\notag\\
&&=0, \label{resonance region}
\end{eqnarray}
where  $\partial/\partial\sigma+v(\partial/\partial\xi)\sim o(v-\lambda)$ and clearly, all the terms in the left hand side of Eq. \eqref{resonance region} are of the same order of magnitude.    We note that the second and the the third terms in Eq. \eqref{resonance region} under the square brackets appear due to the  Landau damping (linear resonance) and particle trapping (nonlinear resonance) respectively. Furthermore,  with the  orderings as above, the quantum effect $\propto R_e$ does not contribute to Eq. \eqref{resonance region} implying that  the results for the trapping will remain qualitatively the same as in the classical theory \cite{vandam1973}. So,   we will focus mainly only on the linear Landau damping.   Thus,  Eq. \eqref{resonance region}, after  the third term being dropped, becomes  
\begin{eqnarray}
&&\epsilon^{-1/2}\left(\frac{m_\alpha}{m_i}\right)^{1/2}\left[\frac{\partial f^{(1)}_\alpha}{\partial\sigma}+v\frac{\partial f^{(1)}_\alpha}{\partial\xi}\right]-\theta_\alpha\left(\frac{m_i}{m_\alpha}\right)\epsilon^{-1}\notag\\
&&\times G_\alpha(v)\frac{\partial\phi^{(1)}}{\partial\xi}=0. \label{equation excluded trapping term}
\end{eqnarray}
Following Ref. \onlinecite{vandam1973} we obtain  
\begin{equation}
\frac{\partial}{\partial\zeta}\sum\theta_\alpha\left<\int_{\text{res.}}f_\alpha dv\right>^{(2)}= c \text{P}\int_{-\infty}^{\infty}\frac{d\zeta'}{\zeta-\zeta'} \frac{\partial\phi^{(1)}(\zeta')}{\partial\zeta'},\label{f_alpha2 in RG}
\end{equation}
where  
\begin{equation}
c=-\epsilon^{-1}\sum\left(\frac{m_i}{m_\alpha}\right) G_\alpha(\lambda).\label{c}
\end{equation}
Substituting Eq. \eqref{f_alpha2 in RG} into Eq. \eqref{modified KdV}, we obtain the following modified KdV equation for IAWs:
\begin{eqnarray}
&&\frac{\partial\phi}{\partial s}+\delta\phi\frac{\partial\phi}{\partial\zeta}+\beta\frac{\partial^3\phi}{\partial\zeta^3}\notag\\
&&+\gamma~\text{P}\int_{-\infty}^{\infty}(\zeta-\zeta')^{-1}\frac{\partial\phi(\zeta')}{\partial\zeta'}d\zeta'=0, \label{KdV}
\end{eqnarray}
where $\phi\equiv\phi^{(1)}$ and the coefficients are given by  $\delta=b/a$, $\beta=1/a$ and $\gamma=c/a$. If the equilibrium distribution of particles   be the Maxwellian [Eq. \eqref{f_alpha Maxwellian distribution}], one can obtain the coefficients $a,~b$ and $c$   (for details see Appendices \ref{appendix-a}-\ref{appendix-c}) as 
\begin{equation}
a=2\lambda^{-2}\left(1+6\lambda^{-2}T^{-1}\right)+\lambda H^2k, \label{a-reduced}
\end{equation}
\begin{equation}
b=3\lambda^{-4}+30T^{-1}\lambda^{-6}-1, \label{b-reduced}
\end{equation}
\begin{equation}
c=\epsilon^{-1}\frac{\lambda}{\sqrt{2\pi}}\left[m^{1/2}+T^{3/2}\text{exp}\left(-\frac{T\lambda^2}{2}\right)\right].\label{c-reduced}
\end{equation}
We note that each of the coefficients $a,~b$ and $c$ are modified by the quantum parameter $H$, in absence of which one recovers the classical results of Vandam \textit{et al.} \cite{vandam1973}. 
\begin{figure*}
\includegraphics[scale=0.5]{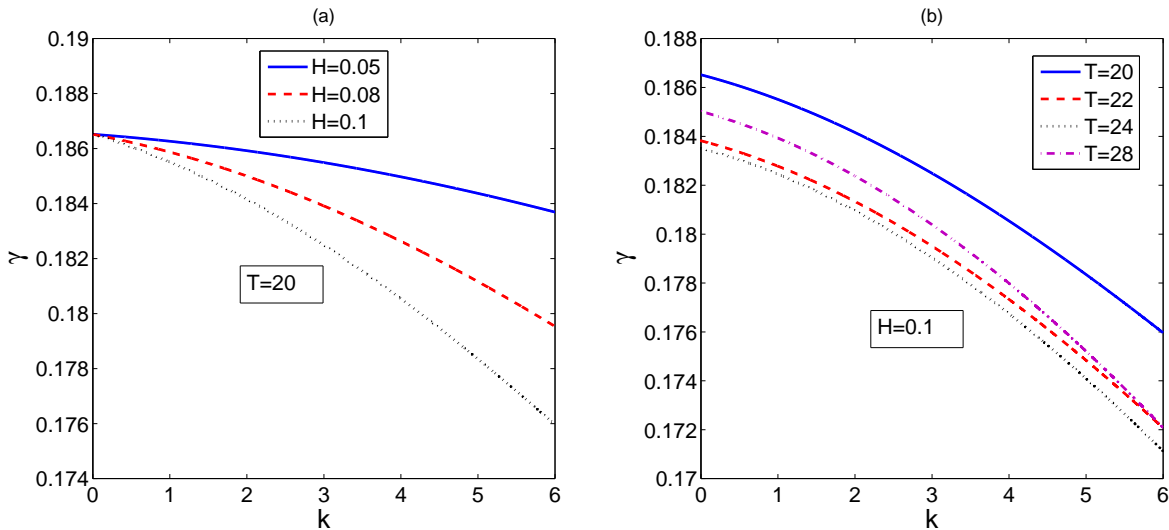}
\caption{The Landau damping rate $\gamma$ (normalized by $\omega_{pi}$) is plotted against the wave number  $k$ (normalized by $\lambda_{D}^{-1}$) in three different cases: (a) when $T$ is fixed and $H$ varies, (b) when $H$ is fixed and $T$ varies and (c) when the value of $T$  is relatively lower than that in the plots (a) and (b).  }
\label{fig:damping}
\end{figure*}
Considering a small effect of the Landau damping ($\varpropto\gamma$) with $\gamma\ll\alpha ~(\gtrsim\beta)$, which holds for $T\gtrsim20$ and $H<1$, we find the solitary wave solution \cite{barman2014} of Eq. \eqref{KdV}  as
\begin{eqnarray}
\phi=&&\Psi~\text{sech}^2\left[\left(\zeta-\frac {\delta}3\int_{0}^{s}\Phi ds\right)/W\right]\notag\\
&&+o(\gamma),\label{soliton-sol}
\end{eqnarray}
where  $\Psi=\Phi_0\left(1+{s}/{s_0}\right)^{-2}$ is the amplitude of the solitary wave solution of the modified KdV equation \eqref{KdV}, and $\Phi=3U_0/\delta$ is the corresponding amplitude,   $W=\left({12\beta}/{\Phi\delta}\right)^{1/2}\equiv\sqrt{4\beta/U_0}$ is the width and $U_0=\Phi\delta/3$ is the constant phase speed (normalized by $c_s$) of the solitary wave solution of the KdV equation in absence of the Landau damping (i.e., when $\gamma$ or $c=0$). For details about the solution, readers are referred to Ref. \onlinecite{barman2014}. Also, $\Phi=\Phi_0$   at $s=0$ and $s_0$ is given by
\begin{equation}
s_0^{-1}=\frac{\gamma}{4}\sqrt{\frac{\delta\Phi_0}{3\beta}}\text{P}\int_{-\infty}^{\infty}\int_{-\infty}^{\infty}\frac{\text{sech}^2z}{z-z'}\frac{\partial}{\partial z'}\left(\text{sech}^2z'\right)dzdz'.\label{tau'}
\end{equation}
\section{Results and discussion}
The profiles of the linear damping rate $\gamma$   are shown in fig. \ref{fig:damping} for different values of $H$ and $T$. From the subplots (a) and (b) it is clear that whatever be the values of $T$ and $H$, the damping rate always decreases with increasing values of the wave number $k$.  The value of $\gamma$ also decreases  with a small increase of   the    quantum parameter $H$, and such a decrement is significant at higher values of $k$. However, subplot (b) shows that there exists a critical value of $T\sim24$, below (above) which the value of $\gamma$ decreases (increases) with increasing value of $T$.     
\par
\begin{figure*}
\includegraphics[scale=0.5]{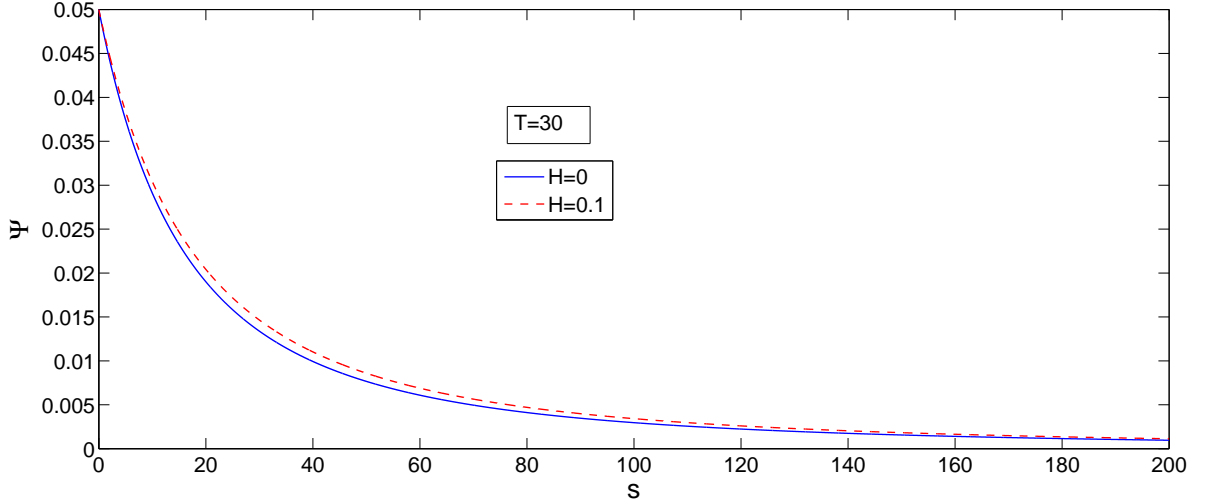}
\caption{The decay of the solitary wave amplitude $\Psi$, given by Eq. \eqref{soliton-sol},  is shown against the space coordinate $s$ for different values of $H$  as in the legend.}
\label{fig:decay-rate}
\end{figure*}

We have also obtained a solitary wave solution \eqref{soliton-sol} of the KdV equation assuming that the Landau damping effect is small compared to those of the  nonlinearity and dispersion. It is found that the wave amplitude decays as time goes on.  The profiles are shown in Fig. \ref{fig:decay-rate} for different values of $H$. It is found that  the decay rate is    lower (than the classical case) in the semiclassical regime  with a  value of $H$. The decay rate remains almost unaltered by the effects of $T$.  
\par
We note that the trapping time of an electron by a solitary pulse is $T_{trap}=\omega_{B}^{-1}$, where $\omega_B$ is the bouncing frequency, given by, $\omega_B= \sqrt{e\phi/m_e}/W$ with $W$ denoting the soliton width. Since  $\phi\sim\epsilon$, $\omega_B\sim\omega_{pe}\sqrt{\epsilon}\sim\omega_{pi}\epsilon^{-1/2}$ (for $\epsilon\sim\sqrt{m_e/m_i}$). However, from Fig. \ref{fig:damping} it is clear that    the Landau damping rate $\gamma$ is of the order of $0.18\omega_{pi}$ (depending on the values of $T,~H$ and $k$), i.e., $\gamma<\omega_B$, the usual criterion for electrons to be trapped. However, in case of ions, since  $\omega_B\sim\omega_{pi}\sqrt{\epsilon}$, we can have $\gamma>\Omega_B$, i.e., ion trapping may be neglected.  Furthermore, one  can compare the magnitudes of the effects of trapping (nonlinear resonance) and the Landau damping (linear resonance). For the nonlinear resonance  we find that \cite{vandam1973}
\begin{eqnarray}
\int_{res.}f_{\alpha} dv\approx&&-\frac{\lambda}{\sqrt{2\pi}}\left[\exp\left(-\frac{1}{2}m\lambda^2\right)\right.\notag\\
&&\left.+\epsilon T^{3/2}\exp\left(-\frac{1}{2}T\lambda^2\right)\right].\label{trap-estimate}
\end{eqnarray}
From Eqs. \eqref{c-reduced} and \eqref{trap-estimate}, it is clear that the effects of the linear resonance is relatively higher than that of the nonlinear one. On the other hand, ions may be reflected  by a solitary pulse and propagate as a precursor, which is also not of interest in the present study.
\par
 In semiclassical plasmas, the thermodynamic temperature of electrons $(T_e)$ is assumed to be larger than the Fermi temperature $(T_F)$ in which case the Pauli blocking is reduced, and the particles' collisional effects can have some role on the dynamics of IAWs. Nevertheless, the inclusion of a collisional term in the semiclassical Vlasov equation is not so straightforward. However, if a small collisional effect (e.g., Coulomb collision) is introduced, the effective electron-electron collision frequency scales as $\nu_{ef}\sim \epsilon\omega_{pe}\left(n_0\lambda_D^3\right)^{-1}\sim\epsilon^{-2}\omega_{pi}\left(n_0\lambda_D^3\right)^{-1}$. For moderate density plasmas with  $n_0\sim6\times10^{23}$ cm$^{-3}$ and $T_e\sim7\times10^6$ K, one can have $\left(n_0\lambda_D^3\right)^{-1}~(\sim0.13)> \epsilon~(\sim0.02)$   and $H\sim0.05$. Thus, $\nu_{ef}~(\gtrsim\epsilon^{-1}\omega_{pi})>\omega_{Be}\sim\sqrt{\epsilon}\omega_{pe}\sim\omega_{pi}\epsilon^{-1/2}$, and consequently, the trapping of electrons will be destroyed. Furthermore, depending on the values of $T$, $H$ and $k$, the Landau damping contribution $c$ can be larger than the damping due to the collisional effects.       
\section{Conclusion} We have studied the effects of   Landau damping associated with the resonance of electrons and ions with ion-acoustic solitary waves in a semiclassical electron-ion plasma. The latter includes the quantum particle dispersion (QPD) in the weak sense, i.e., when    the typical  ion-acoustic    length scale  is larger than the thermal de Broglie wavelength.  A modified KdV equation, which governs the evolution of IAWs, is derived by a multi-scale asymptotic  expansion method. It is found that in contrast to classical results \cite{vandam1973}, the IAW speed $(\lambda)$ and the Landau damping rate $(\gamma)$ are no longer constants by the effects of QPD, but can vary  with the wave number $k$.  Both $\lambda$ and $\gamma$ are seen to be significantly modified by the quantum parameter $H$ and the temperature ratio $T$. It is also found that the decay of the solitary wave amplitude can be suppressed by increasing  the value of $H$. 
\par
To conclude, the nonlinear resonance effects such as trapping, reflection of electrons and ions  should be properly considered especially in the semiclassical and quantum kinetic models in order to describe correctly the  behaviors of IAWs in plasmas.  The dynamics of solitary waves in the (strong) quantum regime, i.e., starting from a Wigner or Wigner-Moyal equation can also be a problem of interest but beyond the scope of the present study. This work is underway and will be communicated elsewhere.    
\acknowledgments{This work was supported by UGC-SAP (DRS, Phase III) with  Sanction  order No.  F.510/3/DRS-III/2015(SAPI),   and UGC-MRP with F. No. 43-539/2014 (SR) and FD Diary No. 3668.}
\appendix
\section{Expressions for the dispersion relation} \label{appendix-dispersion}
Here, we give the details of the derivation of Eq. \eqref{lambda1}.
\begin{equation}
 \sum_\alpha\int_\text{n.r.}(v-\lambda)^{-1}\left(\frac{m_i}{m_{\alpha}}G_{\alpha}(v)+\theta_{\alpha} k^2R\frac{\partial^2G_{\alpha}(v)}{\partial v^2}\right)dv=0. \label{dispersion relation_appendix}
\end{equation}
First of all, we calculate the classical part.
\begin{eqnarray}
&&\sum_\alpha\int_\text{n.r.}(v-\lambda)^{-1}\left(\frac{m_i}{m_\alpha}\right)G_\alpha(v)dv\notag\\
&&=-\frac{1}{\sqrt{2\pi}}\sum_\alpha\left(\frac{m_\alpha}{m_i}\right)^{1/2}\left(\frac{T_e}{T_\alpha}\right)^{3/2}\int_{-\infty}^{\infty}v(v-\lambda)\notag\\
&&\times\text{exp}\left[-\frac{m_\alpha T_e}{m_i T_\alpha}\frac{v^2}{2}\right]dv.\label{calculation for classical lambda}
\end{eqnarray}
Using the transformation $(v^2/2)(m_\alpha T_e/m_i T_\alpha)\longrightarrow v^2$, Eq. \eqref{calculation for classical lambda} yields
\begin{eqnarray}
&&\sum_\alpha\int_\text{n.r.}(v-\lambda)^{-1}\left(\frac{m_i}{m_\alpha}\right)G_\alpha(v)dv\notag\\
&&=-\frac{1}{\sqrt{\pi}}\sum_\alpha\frac{T_e}{T_\alpha}\int_{-\infty}^{\infty}v~\text{exp}(-v^2)\notag\\
&&\times\left(v-\frac{\lambda}{\sqrt{2\frac{m_iT_\alpha}{m_\alpha T_e}}}\right)^{-1}dv. \label{transformed classical lambda}
\end{eqnarray}
Differentiating the Plasma dispersion function once with respect to $\xi$, we obtain
\begin{equation}
Z'(\xi)=\frac{1}{\sqrt{\pi}}\int_{-\infty}^{\infty}\frac{\text{exp}(-t^2)}{(t-\xi)^2}dt. \label{once differentiation of Z_xi, integral form_classical}
\end{equation}
Equation \eqref{once differentiation of Z_xi, integral form_classical}, after  integrating by parts, gives
\begin{equation}
Z'(\xi)=-\frac{1}{\sqrt{\pi}}\int_{-\infty}^{\infty}\frac{2t~\text{exp}(-t^2)}{(t-\xi)}dt=-2[1+\xi Z(\xi)].\label{relation between integral form and differential form_classical lambda}
\end{equation}
Next, using the relation \eqref{relation between integral form and differential form_classical lambda}, we obtain from Eq. \eqref{transformed classical lambda} as
\begin{eqnarray}
&&\sum_\alpha\int_\text{n.r.}(v-\lambda)^{-1}\left(\frac{m_i}{m_\alpha}\right)G_\alpha(v)dv\notag\\
&&=-\sum_\alpha\frac{T_e}{T_\alpha}\left[1+\sqrt{\frac{m_\alpha T_e}{m_i T_\alpha}}\frac{\lambda}{\sqrt{2}}Z\left(\frac{\lambda}{\sqrt{2}}\sqrt{\frac{m_\alpha T_e}{m_i T_\alpha}}\right)\right].\label{classical lambda in terms of Z}
\end{eqnarray} 
Assuming $\lambda\sqrt{m_e/m_i}/\sqrt{2}\ll 1$ and $\lambda\sqrt{T_e/T_i}/\sqrt{2}\gg 1$, we expand the Plasma dispersion function for small as well as large argument \cite{summers1991}, and obtain from Eq. \eqref{classical lambda in terms of Z} as
\begin{eqnarray}
&&\sum_\alpha\int_\text{n.r.}(v-\lambda)^{-1}\left(\frac{m_i}{m_\alpha}\right)G_\alpha(v)dv\notag\\
&&=-(1+T)-\frac{\lambda}{\sqrt{2}}\left[\sqrt{m}(\text{i}\sqrt{\pi}-\lambda\sqrt{2m}-\frac{\text{i}\sqrt{\pi}}{2}m\lambda^2\right.\notag\\
&&\left.+\frac{\sqrt{2}}{3}m^{3/2}\lambda^3)+T^{3/2}(\text{i}\sqrt{\pi}-\frac{\sqrt{2}}{\lambda}T^{-1/2}\right.\notag\\
&&\left.-\frac{\sqrt{2}}{\lambda^3}T^{-3/2}-\frac{3\sqrt{2}}{\lambda^5}T^{-5/2})\right], \label{big classical lambda}
\end{eqnarray}
where, $m=m_e/m_i$ and $T=T_e/T_i$. Neglecting the imaginary terms in the right hand side of Eq. \eqref{big classical lambda}, we obtain 
\begin{eqnarray}
&&\sum_\alpha\int_\text{n.r.}(v-\lambda)^{-1}\left(\frac{m_i}{m_\alpha}\right)G_\alpha(v)dv\notag\\
&&=-1+m\lambda^2\left(1-\frac{m\lambda^2}{3}\right)+\frac{1}{\lambda^2}\left(1+\frac{3}{T\lambda^2}\right). \label{final classical lambda}
\end{eqnarray}
Next, we calculate the contribution from the quantum particle dispersion.
\begin{eqnarray}
&&-R_ek^2\int_\text{n.r.}(v-\lambda)^{-1}\frac{\partial^2G_e(v)}{\partial v^2}dv\notag\\
&&=\frac{H^2k^2}{24}\left(\frac{m_i}{m_e}\right)^2\left[(v-\lambda)^{-1}\frac{\partial G_e(v)}{\partial v}+\int(v-\lambda)^{-2}\right.\notag\\
&&\left.\times\frac{\partial G_e(v)}{\partial v} dv\right]_{\text{n.r.}}\notag\\
&&=\frac{H^2k^2}{12}\frac{1}{\sqrt{2\pi}}\left(\frac{m_i}{m_e}\right)^{1/2}\int_{-\infty}^{\infty}v(v-\lambda)^{-3}\notag\\
&&\times\text{exp}\left[-\frac{m_e}{m_i}\frac{v^2}{2}\right]dv.\label{calculation for quantum lambda}
\end{eqnarray}
Using the transformation $(v^2/2)(m_e/m_i)\longrightarrow v^2$, Eq. \eqref{calculation for quantum lambda} yields
\begin{eqnarray}
&&-R_ek^2\int_\text{n.r.}(v-\lambda)^{-1}\frac{\partial^2G_e(v)}{\partial v^2}dv\notag\\
&&=\frac{H^2k^2}{24}\frac{1}{\sqrt{\pi}}\int_{-\infty}^{\infty}v~\text{exp}(-v^2)\notag\\
&&\times\left[v-\frac{\lambda}{\sqrt{2\frac{m_i}{m_e}}}\right]^{-3}dv. \label{transformed quantum lambda}
\end{eqnarray}
Next, differentiating the Plasma dispersion function  thrice with respect to $\xi$, we obtain
\begin{equation}
Z'''(\xi)=-4[(2\xi^2-3)\xi Z(\xi)+2\xi^2-2]. \label{thrice differentiation of Z_xi_quantum}
\end{equation}
Also, 
\begin{equation}
Z'''(\xi)=-\frac{4}{\sqrt{\pi}}\int_{-\infty}^{\infty}\frac{t~\text{exp}(-t^2)}{(t-\xi)^3}dt, \label{thrice differentiation of Z_xi, integral form_quantum}
\end{equation}
From Eqs. \eqref{thrice differentiation of Z_xi_quantum} and \eqref{thrice differentiation of Z_xi, integral form_quantum}, we have 
\begin{equation}
(2\xi^2-3)\xi Z(\xi)+2\xi^2-2=\frac{1}{\sqrt{\pi}}\int_{-\infty}^{\infty}\frac{t~\text{exp}(-t^2)}{(t-\xi)^3}dt.\label{relation between integral form and differential form_quantum lambda}
\end{equation}
Next, using the relation \eqref{relation between integral form and differential form_quantum lambda}, we obtain from Eq. \eqref{transformed quantum lambda} as
\begin{eqnarray}
&&-R_ek^2\int_\text{n.r.}(v-\lambda)^{-1}\frac{\partial^2G_e(v)}{\partial v^2}dv\notag\\
&&=\frac{H^2k^2}{24}\left[\left(\lambda^2\frac{m_e}{m_i}-3\right)\frac{\lambda}{\sqrt{2}}\sqrt{\frac{m_e}{m_i}}\right.\notag\\
&&Z\left.\left(\frac{\lambda}{\sqrt{2}}\sqrt{\frac{m_e}{m_i}}\right)+\frac{m_e}{m_i}\lambda^2-2\right].\label{quantum lambda in terms of Z}
\end{eqnarray} 
Assuming $\lambda\sqrt{m_e/m_i}/\sqrt{2}\ll 1$, and expanding the Plasma dispersion function for small argument \cite{summers1991} upto $O(\lambda^3)$, we obtain from Eq. \eqref{quantum lambda in terms of Z} as
\begin{eqnarray}
&&-Rk^2\int_\text{n.r.}(v-\lambda)^{-1}\frac{\partial^2G_e(v)}{\partial v^2}dv\notag\\
&&=\frac{H_e^2k^2}{24}\left[(m\lambda^2-3)\frac{\lambda}{\sqrt{2}}\sqrt{m}(\text{i}\sqrt{\pi}-\lambda\sqrt{2m}-\frac{\text{i}\sqrt{\pi}}{2}m\lambda^2\right.\notag\\
&&\left.+\frac{\sqrt{2}}{3}m^{3/2}\lambda^3)+m\lambda^2-2\right]. \label{big quantum lambda}
\end{eqnarray}
Neglecting the imaginary terms in the right hand side of Eq. \eqref{big quantum lambda}, we obtain 
\begin{eqnarray}
&&-R_ek^2\int_\text{n.r.}(v-\lambda)^{-1}\frac{\partial^2G_e(v)}{\partial v^2}dv\notag\\
&&=\frac{H^2k^2}{24}m\left[\frac13m^2\lambda^6+4\lambda^2-\frac{2}{m}\right]\notag\\
&&=\frac{H^2k^2}{24}\left[\frac13m^3\lambda^6+4m\lambda^2-2\right]. \label{final quantum lambda}
\end{eqnarray}
Next, substituting the values from Eqs. \eqref{final classical lambda} and \eqref{final quantum lambda} into Eq. \eqref{dispersion relation_appendix}, we obtain
\begin{eqnarray}
&&-1+m\lambda^2-\frac13m^2\lambda^4+\frac{1}{\lambda^2}+\frac{3}{T}\frac{1}{\lambda^4}+\frac{H^2k^2}{24}\left(\frac13m^3\lambda^6\right.\notag\\
&&\left.+4m\lambda^2-2\right)=0, \label{classical+quantum_lambda}
\end{eqnarray}
Neglecting the smaller terms compared to the larger ones in Eq. \eqref{classical+quantum_lambda}, we obtain
\begin{equation}
\left(1+\frac{H^2k^2}{12}\right)\lambda^4-\lambda^2-\frac{3}{T}=0.\label{classical+quantum_DR}
\end{equation}
Solving Eq. \eqref{classical+quantum_DR}, we obtain
\begin{equation}
\lambda^2=\frac{1\pm\sqrt{1+\frac{12}{T}\left(1+\frac{H^2k^2}{12}\right)}}{2\left(1+\frac{H^2k^2}{12}\right)}.\label{lambda^2}
\end{equation}
Neglecting the smaller terms, we obtain the expression for nonlinear wave speed from Eq.\eqref{lambda^2} as
\begin{equation}
\lambda=1+\frac{3}{2T}-\frac{H^2k^2}{24}. \label{final lambda}
\end{equation}
\section{The coefficient $a$} \label{appendix-a}
We   simplify  the expression for  $a$ [Eq. \eqref{a}] as 
\begin{eqnarray}
a&&=-\left[\lambda\sum_\alpha\left(\frac{m_i}{m_\alpha}\right)\int_{\text{n.r.}}(v-\lambda)^{-1}\frac{\partial G_\alpha(v)}{\partial v}dv\right.\notag\\
&&\left.+18k\sum_\alpha\theta_\alpha R_\alpha\left(\frac{m_i}{m_\alpha}\right)\int_{\text{n.r.}}(v-\lambda)^{-4}G_\alpha(v)dv\right], \label{appendix a}
\end{eqnarray}
The first term on the right hand side of Eq. \eqref{appendix a} is
\begin{eqnarray}
&&-\lambda\sum_\alpha\left(\frac{m_i}{m_\alpha}\right)\int_{\text{n.r}}(v-\lambda)^{-1}\frac{\partial G_\alpha(v)}{\partial v}dv\notag\\
=&&-\lambda\sum_\alpha\left(\frac{m_i}{m_\alpha}\right)\left[(v-\lambda)^{-1}G_\alpha(v)+\int\frac{G_\alpha(v)}{(v-\lambda)^{2}} dv\right]_{\text{n.r.}}\notag\\
=&&-\lambda\sum_\alpha\left(\frac{m_i}{m_\alpha}\right)\int_{\text{n.r}}(v-\lambda)^{-2}G_\alpha(v)dv,\notag\\
=&&\lambda\sum_\alpha\left(\frac{T_e}{T_\alpha}\right)\int_{-\infty}^{\infty}v(v-\lambda)^{-2}f^{(0)}_\alpha(v)dv\notag\\
=&&\frac{\lambda}{\sqrt{2\pi}}\sum_\alpha\frac{T_e}{T_\alpha}\sqrt{\frac{m_\alpha T_e}{m_iT_\alpha}}\int_{-\infty}^{\infty}\frac{v}{(v-\lambda)^{-2}}\text{exp}\notag\\
&&\times\left[-\frac{m_\alpha T_e}{m_iT_\alpha}\frac{v^2}{2}\right]dv.\label{calculation of a}
\end{eqnarray}
Differentiating the Plasma dispersion function  twice with respect to $\xi$, we obtain
\begin{equation}
Z''(\xi)=-2[(1-2\xi^2)Z(\xi)-2\xi]. \label{twice differentiation of Z_xi}
\end{equation}
Also, 
\begin{equation}
Z''(\xi)=-\frac{2}{\sqrt{\pi}}\int_{-\infty}^{\infty}\frac{t~\text{exp}(-t^2)}{(t-\xi)^2}dt, \label{twice differentiation of Z_xi, integral form}
\end{equation}
Equations \eqref{twice differentiation of Z_xi} and \eqref{twice differentiation of Z_xi, integral form} yield
\begin{equation}
(1-2\xi^2)Z(\xi)-2\xi=\frac{1}{\sqrt{\pi}}\int_{-\infty}^{\infty}\frac{t~\text{exp}(-t^2)}{(t-\xi)^2}dt.\label{relation between integral form and differential form_a}
\end{equation}
Using the transformation $(v^2/2)(m_\alpha T_e/m_i T_\alpha)\longrightarrow v^2$, we see that Eq. \eqref{calculation of a} can be rewritten as
\begin{eqnarray}
&&-\lambda\sum_\alpha\left(\frac{m_i}{m_\alpha}\right)\int_{\text{n.r}}(v-\lambda)^{-1}\frac{\partial G_\alpha(v)}{\partial v}dv\notag\\
=&&\frac{\lambda}{\sqrt{2\pi}}\sum_\alpha\frac{T_e}{T_\alpha}\sqrt{\frac{m_\alpha T_e}{m_iT_\alpha}}\int_{-\infty}^{\infty}v~\text{exp}(-v^2)\notag\\
&&\times\left(v-\frac{\lambda}{\sqrt{2\frac{m_i T_\alpha}{m_\alpha T_e}}}\right)dv.\label{transformed a}
\end{eqnarray}
Using the relation \eqref{relation between integral form and differential form_a}, Eq. \eqref{transformed a} reduces to
\begin{eqnarray}
&&-\lambda\sum_\alpha\left(\frac{m_i}{m_\alpha}\right)\int_{\text{n.r}}(v-\lambda)^{-1}\frac{\partial G_\alpha(v)}{\partial v}dv\notag\\
=&&\frac{\lambda}{\sqrt{2}}\sum_\alpha\frac{T_e}{T_\alpha}\sqrt{\frac{m_\alpha T_e}{m_iT_\alpha}}\left[\left(1-\lambda^2\frac{m_\alpha T_e}{m_i T_\alpha}\right)Z\left(\frac{\lambda}{\sqrt{2}}\sqrt{\frac{m_\alpha T_e}{m_i T_\alpha}}\right)\right.\notag\\
&&\left.-\lambda\sqrt{2\frac{m_\alpha T_e}{m_i T_\alpha}}\right]. \label{a in terms of Z}
\end{eqnarray}
Assuming $\lambda\sqrt{m_e/m_i}/\sqrt{2}\ll 1$ and $\lambda\sqrt{T_e/T_i}/\sqrt{2}\gg 1$, and expanding the Plasma dispersion function for small argument upto $o(\lambda^5)$ as well as for large argument upto $O(\lambda^{-7})$ \cite{summers1991}, we obtain from Eq. \eqref{a in terms of Z} the following
\begin{eqnarray}
&&-\lambda\sum_\alpha\left(\frac{m_i}{m_\alpha}\right)\int_{\text{n.r}}(v-\lambda)^{-1}\frac{\partial G_\alpha(v)}{\partial v}dv\notag\\
=&&\frac{\lambda}{\sqrt{2}}\left[\sqrt{m}[(1-m\lambda^2)(\text{i}\sqrt{\pi}-\lambda\sqrt{2m}-\frac{\text{i}\sqrt{\pi}}{2}m\lambda^2\right.\notag\\
&&\left.+\frac{\sqrt{2}}{3}m^{3/2}\lambda^3+\frac{\text{i}\sqrt{\pi}}{8}m^2\lambda^4-\frac{\sqrt{2}}{15}m^{5/2}\lambda^5)\right.\notag\\
&&\left.-\sqrt{2m}\lambda]+T^{3/2}[(1-T\lambda^2)(\text{i}\sqrt{\pi}-\frac{\sqrt{2}}{\lambda}T^{-1/2}\right.\notag\\
&&\left.-\frac{\sqrt{2}}{\lambda^3}T^{-3/2}-\frac{3\sqrt{2}}{\lambda^5}T^{-5/2}-\frac{105\sqrt{2}}{\lambda^9}T^{-9/2}\right.\notag\\
&&\left.-\frac{15\sqrt{2}}{\lambda^7}T^{-7/2})-\sqrt{2T}\lambda]\right], \label{big a}
\end{eqnarray}
Neglecting the imaginary part, Eq. \eqref{big a} reduces to
\begin{eqnarray}
&&-\lambda\sum_\alpha\left(\frac{m_i}{m_\alpha}\right)\int_{\text{n.r}}(v-\lambda)^{-1}\frac{\partial G_\alpha(v)}{\partial v}dv\notag\\
=&&\frac{\lambda}{\sqrt{2}}\left(-2\sqrt{2}m^{3/2}\lambda+\frac{4\sqrt{2}}{3}m^{2}\lambda^3-\frac{2\sqrt{2}}{5}m^{3}\lambda^5\right.\notag\\
&&\left.-\frac{\sqrt{2}}{15}m^{4}\lambda^7+\frac{2\sqrt{2}}{\lambda^3}+\frac{12\sqrt{2}}{\lambda^5}\frac{1}{T}+\frac{90\sqrt{2}}{\lambda^7}\frac{1}{T^2}\right.\notag\\
&&\left.-\frac{105\sqrt{2}}{\lambda^9}\frac{1}{T^3}\right).\label{real part big a1}
\end{eqnarray}
In a similar procedure, we obtain the second term on the right hand side of Eq. \eqref{appendix a} as
\begin{eqnarray}
&&-18k\sum_\alpha\theta_\alpha R_\alpha\left(\frac{m_i}{m_\alpha}\right)\int_{\text{n.r.}}(v-\lambda)^{-4}G_\alpha(v)dv\notag\\
=&&\frac{H_e^2k}{8}\left(8\lambda-8\lambda^3m+\frac{48}{15}\lambda^5m^2-\frac{11}{15}\lambda^7m^3\right.\notag\\
&&\left.+\frac{1}{15}\lambda^9m^4\right), \label{real part big a2}
\end{eqnarray}
Next, adding Eqs. \eqref{real part big a1} and \eqref{real part big a2}, and  neglecting the smaller terms in comparison with the larger ones, we obtain
\begin{eqnarray}
a=&&\lambda^{-2}(2+12\lambda^{-2}T^{-1}+90\lambda^{-4}T^{-2}\notag\\
&&-105\lambda^{-6}T^{-3})+\lambda H_e^2k. \label{final a}
\end{eqnarray}
\section{The coefficient $b$} \label{appendix-b}
We simplify the expression for  $b$ [Eq. \eqref{b}] as
\begin{eqnarray}
b=&&\sum_\alpha\theta_\alpha\left(\frac{m_i}{m_\alpha}\right)\int_{\text{n.r}}(v-\lambda)^{-3}G_\alpha(v)dv\notag\\
=&&-\sum\theta_\alpha\left(\frac{T_e}{T_\alpha}\right)\int_{-\infty}^{\infty}v(v-\lambda)^{-3}f^{(0)}_\alpha(v)dv\notag\\
=&&-\frac{1}{\sqrt{2\pi}}\sum_\alpha\theta_\alpha\frac{T_e}{T_\alpha}\sqrt{\frac{m_\alpha T_e}{m_iT_\alpha}}\int_{-\infty}^{\infty}\frac{v}{(v-\lambda)^{-3}}~\text{exp}\notag\\
&&\times\left[-\frac{m_\alpha T_e}{m_iT_\alpha}\frac{v^2}{2}\right]dv.\label{calculation of b}
\end{eqnarray}
Differentiating the Plasma dispersion function  thrice with respect to $\xi$, and noting that
\begin{equation}
Z'''(\xi)=-\frac{4}{\sqrt{\pi}}\int_{-\infty}^{\infty}\frac{t~\text{exp}(-t^2)}{(t-\xi)^3}dt, \label{thrice differentiation of Z_xi, integral form}
\end{equation}
 we obtain
\begin{equation}
(2\xi^2-3)\xi Z(\xi)+2\xi^2-2=\frac{1}{\sqrt{\pi}}\int_{-\infty}^{\infty}\frac{t~\text{exp}(-t^2)}{(t-\xi)^3}dt.\label{relation between integral form and differential form_b}
\end{equation}
Using the transformation $(v^2/2)(m_\alpha T_e/m_i T_\alpha)\longrightarrow v^2$, we see that Eq. \eqref{calculation of b} can be rewritten as
\begin{eqnarray}
b=&&-\frac{1}{2\sqrt{\pi}}\sum_\alpha\theta_\alpha\frac{T_e^2m_\alpha}{m_iT_iT_\alpha}\int_{-\infty}^{\infty}v~\text{exp}(-v^2)\notag\\
&&\times\left(v-\frac{\lambda}{\sqrt{2\frac{m_i T_\alpha}{m_\alpha T_e}}}\right)^3dv.\label{transformed b}
\end{eqnarray}
Using the relation \eqref{relation between integral form and differential form_b}, Eq. \eqref{transformed b} becomes
\begin{eqnarray}
b=&&-\frac12\sum_\alpha\theta_\alpha\frac{T_e^2m_\alpha}{m_iT_iT_\alpha}\left[\left(\lambda^2\frac{m_\alpha T_e}{m_i T_\alpha}-3\right)\frac{\lambda}{\sqrt{2}}\sqrt{\frac{m_\alpha T_e}{m_i T_\alpha}}\right.\notag\\ &&Z\left.\left(\frac{\lambda}{\sqrt{2}}\sqrt{\frac{m_\alpha T_e}{m_i T_\alpha}}\right)+\frac{m_\alpha T_e}{m_i T_\alpha}\lambda^2-2\right]. \label{b in terms of Z}
\end{eqnarray}
Assuming $\lambda\sqrt{m_e/m_i}/\sqrt{2}\ll 1$ and $\lambda\sqrt{T_e/T_i}/\sqrt{2}\gg 1$ and expanding the Plasma dispersion function for small argument upto $o(\lambda^5)$ as well as for large argument upto $o(\lambda^{-7})$  \cite{summers1991}, we obtain from Eq. \eqref{b in terms of Z} the following
\begin{eqnarray}
b=&&\frac12\left[-\frac{m_eT_e}{m_iT_i}[(3-m\lambda^2)\frac{\lambda}{\sqrt{2}}\sqrt{m}(\text{i}\sqrt{\pi}-\lambda\sqrt{2m}\right.\notag\\
&&\left.-\frac{\text{i}\sqrt{\pi}}{2}m\lambda^2+\frac{\sqrt{2}}{3}m^{3/2}\lambda^3+\frac{\text{i}\sqrt{\pi}}{8}m^2\lambda^4-\frac{\sqrt{2}}{15}m^{5/2}\lambda^5)\right.\notag\\
&&\left.-m\lambda^2+2]+T^2[(3-T\lambda^2)\frac{\lambda}{\sqrt{2}}\sqrt{T}(\text{i}\sqrt{\pi}-\frac{\sqrt{2}}{\lambda}T^{-1/2}\right.\notag\\
&&\left.-\frac{\sqrt{2}}{\lambda^3}T^{-3/2}-\frac{3\sqrt{2}}{\lambda^5}T^{-5/2}-\frac{105\sqrt{2}}{\lambda^9}T^{-9/2}\right.\notag\\
&&\left.-\frac{15\sqrt{2}}{\lambda^7}T^{-7/2})-T\lambda^2+2\lambda]\right], \label{big b}
\end{eqnarray}
Next, considering the real parts of $b$,  Eq. \eqref{big b} gives
\begin{eqnarray}
b=&&\frac{1}{30}[mT(60m\lambda^2-30m^2\lambda^4+6m^3\lambda^6-m^4\lambda^8-30)\notag\\
&&+900T^{-1}\lambda^{-6}-4725T^{-2}\lambda^{-8}+90\lambda^{-4}]. \label{real part big b}
\end{eqnarray}
Neglecting the small terms and also using the fact that, $(T_e/T_i)(m_e/m_i)\sim o(1)$, we have from Eq. \eqref{real part big b}
\begin{eqnarray}
b&&\simeq\frac{1}{30}[-30+900T^{-1}\lambda^{-6}+90\lambda^{-4}]\notag\\
&&=3\lambda^{-4}-1+30T^{-1}\lambda^{-6}. \label{final b}
\end{eqnarray}
\section{The coefficient $c$} \label{appendix-c}
We simplify the expression for $c$ [Eq. \eqref{c}] as
\begin{eqnarray}
c=&&-\epsilon^{-1}\sum_\alpha\left(\frac{m_i}{m_\alpha}\right) G_\alpha(\lambda)\notag\\
=&&\epsilon^{-1}\frac{\lambda}{\sqrt{2\pi}}\left[m^{1/2}\text{exp}\left(-\frac{m\lambda^2}{2}\right)\right.\notag\\
&&\left.+T^{3/2}\text{exp}\left(-\frac{T\lambda^2}{2}\right)\right],\label{c_appendix}
\end{eqnarray}
where, we have used the expression for $G_\alpha(\lambda)$ as
\begin{equation}
G_\alpha(\lambda)=-\frac{\lambda}{\sqrt{2\pi}}\left(\frac{m_\alpha T_e}{m_iT_\alpha}\right)^{3/2}\text{exp}\left[-\frac{m_\alpha T_e}{m_i T_\alpha}\frac{\lambda^2}{2}\right],\label{G_alpha(lambda)}
\end{equation}
Assuming $\text{exp}(-m\lambda^2/2)\approx1$, we obtain from Eq. \eqref{c_appendix} the following expression for $c$
\begin{eqnarray}
c=&&\epsilon^{-1}\frac{\lambda}{\sqrt{2\pi}}\left[m^{1/2}+T^{3/2}\text{exp}\left(-\frac{T\lambda^2}{2}\right)\right].\label{final c}
\end{eqnarray}

\end{document}